\documentclass[8.5pt,twoside,twocolumn]{article}
\usepackage[colorlinks=true,urlcolor=blue]{hyperref} 
\usepackage[super,sort&compress,comma]{natbib} 
\usepackage[version=3]{mhchem} 
\usepackage{times,mathptmx}      
\usepackage{sectsty}
\usepackage{balance} 
\usepackage{graphicx} 
\usepackage{lastpage}
\usepackage[format=plain,justification=raggedright,singlelinecheck=false,font=small,labelfont=bf,labelsep=space]{caption} 
\usepackage{fancyhdr}
\usepackage{array}
\usepackage[figuresright]{rotating} 
\usepackage{color}
\usepackage{amsmath}
\usepackage{amssymb}
\usepackage{MnSymbol}
\newcommand{\un}[1]{\ensuremath{\unskip\,\mathrm{#1}}}

\DeclareGraphicsExtensions{.eps}

\begin{document}

\pagestyle{plain}
\renewcommand{\thefootnote}{\fnsymbol{footnote}}
\setlength{\arrayrulewidth}{1pt}
\setlength{\columnsep}{6.5mm}
\setlength\bibsep{1pt}

\twocolumn[
  \begin{@twocolumnfalse}
\noindent\LARGE{\textbf{Morphology of gold nanoparticles determined by full-curve fitting of the light absorption spectrum. Comparison with X-ray scattering and electron microscopy data$^\dag$}}
\vspace{0.6cm}

\noindent\large{\textbf{Kostyantyn Slyusarenko,$^{\ast}$ Benjamin Ab\'{e}cassis, Patrick Davidson, and Doru Constantin$^{\ast}$}}\vspace{0.2cm}

\noindent \textit{Laboratoire de Physique des Solides, Univ. Paris-Sud, CNRS, UMR8502, 91405 Orsay Cedex, France.}\\ \href{mailto:kslyusarenko@gmail.com}{kslyusarenko@gmail.com}, \href{mailto:doru.constantin@u-psud.fr}{doru.constantin@u-psud.fr}

\vspace{0.6cm}
\noindent Published in \textit{Nanoscale}, \textbf{6}(22), 13527-13534 (2014).

\noindent \textbf{DOI:\href{http://dx.doi.org/10.1039/c4nr04155k}{10.1039/c4nr04155k}}
\vspace{0.6cm}

\noindent \normalsize{UV-Vis absorption spectroscopy is frequently used to characterize the size and shape of gold nanoparticles. We present a full-spectrum model that yields reliable results for the commonly encountered case of mixtures of spheres and rods in varying proportions. We determine the volume fractions of the two populations, the aspect ratio distribution of the nanorods (average value and variance) and the interface damping parameter. We validate the model by checking the fit results against small-angle X-ray scattering and transmission electron microscopy data and show that correctly accounting for the polydispersity in aspect ratio is essential for a quantitative description of the longitudinal plasmon peak.}
\vspace{0.6cm}

\centerline{\includegraphics[width=9cm,angle=0]{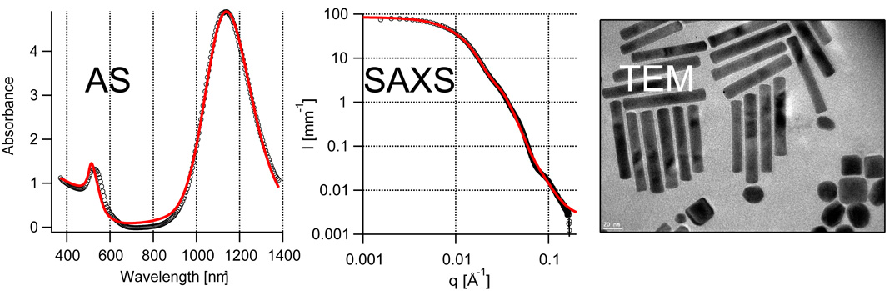}}

\vspace{0.6cm}
 \end{@twocolumnfalse}
  ]
 
\footnotetext{\dag~Electronic Supplementary Information (ESI) available: TEM images of nanoparticles and detailed analysis, simplified relations for the AS model, alternative estimate for the concentration and discussion of the dielectric constant chosen for bulk gold. See the end of this file.}

\section{Introduction}
\label{sec:intro}

\nocite{Eustis:2006b} 
The importance of gold nanoparticles in physics, biology and materials science over the last decade can hardly be overstated. Their manifold applications (see Ref.~\citenum{Eustis:2006b} and references therein) motivated a sustained effort \cite{Daniel:2004,Grzelczak:2008} towards the reproducible synthesis of particles (more commonly, spheres and rods) with controlled shape and size. Once produced, the particles must be characterized as quickly and as thoroughly as possible. In general, this is first done by UV-Vis-IR absorption spectroscopy (AS), a widely available technique that provides valuable information, since the absorption spectrum of noble metal nanoparticles is extremely sensitive to their shape (though much less to their absolute size).

The optical properties of metal nanoparticles have been extensively studied \cite{Kreibig:1995}. For particles much smaller than the wavelength, in the electrostatic approximation, the simplest way of describing the AS data is by the Gans theory \cite{Gans12}, with a finite-size correction to the dielectric function of the metal \cite{Genzel:1975}. This approach was widely used for noble metal nanospheres \cite{Mulvaney:1996,Link:1999,Amendola:2009} and nanorods \cite{Link:1999b,Brioude:2005}. In general, the authors only compare the experimental plasmon peak position with the simulation result, but some of the models also describe its width, by including the effect of  polydispersity, for the sphere radius \cite{Amendola:2009} and for the aspect ratio of rods \cite{Eustis:2006}. With many synthesis conditions, the final state consists of rod and sphere mixtures, each population being polydisperse in size and (for the rods) in aspect ratio \cite{Hubert:2010}. Any realistic model should take these complications into account.

The analysis is usually complemented by transmission electron microscopy (TEM), a technique with the major advantages of imaging the particles directly and of requiring no assumptions as to their environment. However, accurate statistical analysis of the TEM images is very work-intensive and relies on the assumption that the distribution of deposited and analysed particles is the same as in the original solution. Moreover, TEM does not give access to the particle concentrations.

\clearpage

\begin{figure*}[t]
\centerline{\includegraphics[width=0.75\textwidth,angle=0]{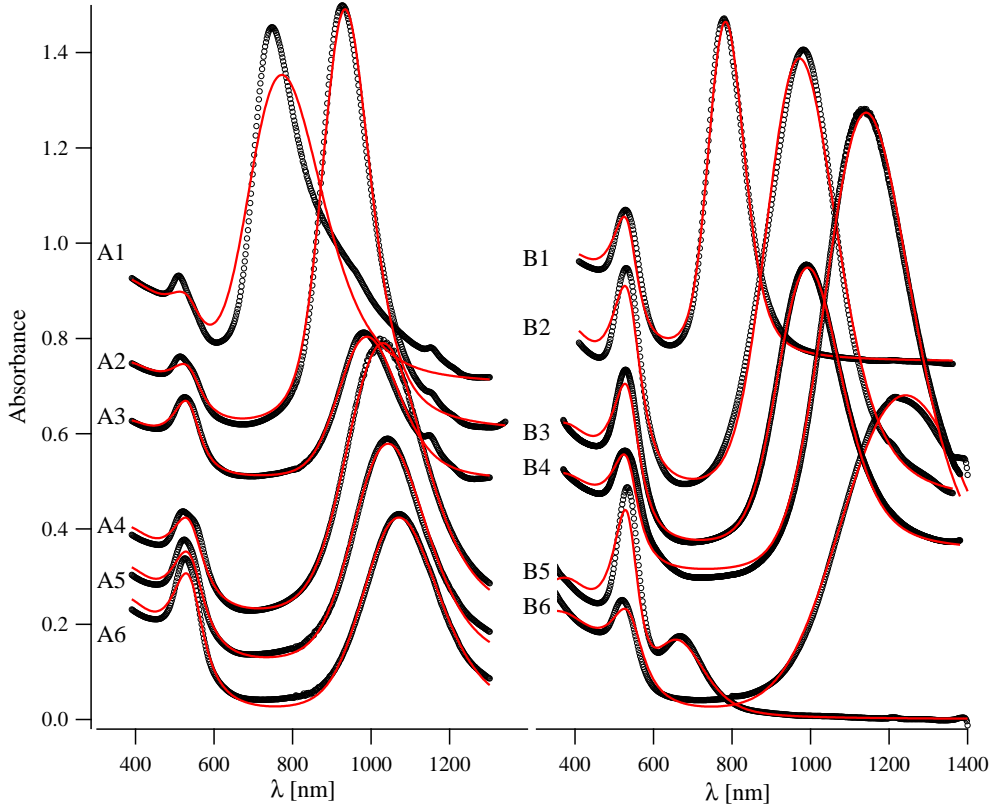}}
\caption{Fits to the absorbance spectroscopy data with the model~\protect{\eqref{eq:Gans}}. The curves were shifted vertically for clarity.}
\label{fig:UV_Vis}
\end{figure*}

Small-angle X-ray scattering (SAXS) does not perturb the solution, yields results averaged over millimetre-size volumes and over times ranging from milliseconds to minutes and is practically insensitive to the organic components (whose contribution is negligible compared to that of the inorganic nanoparticles). The modelling process makes no particular assumptions beyond solution homogeneity and is very sensitive to the particle size, but when different types of particles are present it is quite difficult to discriminate between them.

In this article, we develop a model that allows fitting the entire AS curve, while taking into account the polydispersity of the aspect ratio. We use it to study solutions obtained in diverse synthesis conditions, yielding mixtures of nanospheres and nanorods with varying radius and aspect ratio. Furthermore, the polydispersities of the two populations and their relative proportion change from sample to sample. This provides an ideal testing ground for the accuracy and discriminating power of the model.

After analyzing the absorption spectra by full-curve fitting we validate the results through a comparison with SAXS and TEM data. More specifically:
\begin{itemize}
\item Fitting the AS curves with reasonable starting values for the radii of the spheres and rods ($R_S = 10 \, \text{nm}$ and $R_R = 5  \, \text{nm}$, respectively) yields the volume fractions of spheres and rods ($\phi_S$ and $\phi_R$) and the aspect ratio $X$ of the rods, defined as the length-to-diameter ratio: $X = L/(2 R_R)$ as well as its polydispersity $\sigma_{X}$.
\item The volume fractions and $X$ are used as starting parameters for fitting the SAXS intensity $I(q)$, which then yields the radii for the spheres and rods and their polydispersities ($\sigma_S$ and $\sigma_R$). $R_S$ and $R_R$ are then injected back into the AS fitting for a refined analysis.
\item Finally, the results are checked against the values derived from TEM images for $(R_S, \sigma_S)$, $(R_R, \sigma_R)$ and $(L, \sigma_L)$.
\end{itemize}

The comparison of the three characterization techniques shows that the full-curve model for the AS data is quite robust and yields accurate values for the average aspect ratio of the rods and its variance, as well as for the particle concentrations. It can thus be employed for rapid evaluation of synthesis protocols and provides detailed information even in the absence of time-consuming techniques such as TEM. For ease of use, we also provide simplified relations between the features of the AS curves, on the one hand, and the morphological parameters of the particles and their concentration, on the other hand.

\section{Methods}
\label{sec:Meth}

\subsection{Gold nanorod synthesis}
\label{subsec:Synth}

The nanoparticles were synthesized by a seed-growth method, in the presence of either sodium oleate \cite{Ye:2013} (series A) or bromosalicylic acid \cite{Ye:2012} (series B); we changed the pH and the amount of seed solution to obtain varied particle characteristics (e.g.\ the aspect ratio of the nanorods). For the seed solution, 5~mL of 0.5~mM \ce{HAuCl4} solution was mixed with 5~mL of 0.2~M hexadecyltrimethylammonium bromide (CTAB) solution. 1~mL of a fresh 6~mM \ce{NaBH4} solution was injected into the \ce{HAuCl4}-CTAB mixture under vigorous stirring at 1200 rpm. The solution color changed from yellow to brownish-yellow, and the stirring was stopped after 2 min. The seed solution was aged at room temperature for 30~min before injection into the growth solution.

The growth solution was prepared by mixing 9.0~g of CTAB either with 1.234~g of sodium oleate (series A) or with 1.1~g of 5-bromosalicylic acid (series B) dissolved in 250~mL of distilled water at 30~$^{\circ}$C. A varying amount of 4~mM \ce{AgNO3} solution was then added (see Table~\ref{table:synth}). The mixture was kept undisturbed at 30~$^{\circ}$C for 15~min, after which 250~mL of a 1~mM \ce{HAuCl4} solution and a small amount of 37~wt.\% \ce{HCl} solution were added (see Table~\ref{table:synth}).

\begin{table}
\begin{small}
\begin{tabular}{|c||c|c|c|c|c|c|}
\hline  Sample  &A1  &A2  &A3  &A4  &A5  &A6 \\ 
\hline  \ce{AgNO3}  &36  &36  &36  &36  &36  &36 \\ 
\hline  \ce{HCl} &2  &2  &3  &4.3  &5  &6 \\ 
\hline 
\hline  Sample  &B1  &B2  &B3  &B4  &B5  &B6  \\ 
\hline  \ce{AgNO3} &24  &24  &36  &24  &12  &30	\\ 
\hline  \ce{HCl}  &0  &2  &2  &2  &3  &3 \\ 
\hline 
\end{tabular}
\end{small}
\caption{Synthesis conditions. Amounts of \ce{AgNO3} and \ce{HCl} solutions added to the growth solution (in mL).}
\label{table:synth}
\end{table}

After slowly stirring at 400~rpm for 15~min, 1.25~mL of 64~mM ascorbic acid solution was added and the mixture was strongly stirred until it became colorless. Finally, 0.4~mL (series A) or 0.8~mL (series B) seed solution was injected into the growth solution. The resulting mixture was stirred for 30~s and left undisturbed at 30~$^{\circ}$C for 12~h, allowing the particles to grow. The particles were washed twice, by centrifuging (at 6000~g for 30~min) and redispersed in 0.1~M CTAB.

\subsection{Absorbance spectroscopy: measurement and analysis}
\label{subsec:Anal}

We used a Cary~5000 spectrometer (Agilent) to measure the absorption spectrum of dilute particle solutions between 400 and 1400~nm. The synthesis solutions were diluted 25 to 100 times (yielding a total Au volume fraction $\phi = \phi_S + \phi_R$ of the order of $10^{-6}$) and held in polystyrene cuvettes with 10~mm optical path.

The absorption spectra of gold nanorods consist of two bands corresponding to the oscillation of the free electrons either parallel or perpendicular to the long axis of the nanorods. The transverse plasmon peak (TPP) has a maximum around 520~nm and depends only weakly on the nanorod diameter. The longitudinal plasmon peak (LPP) is much stronger than the TPP and shifts to larger wavelength as the aspect ratio $X$ increases. To describe the spectrum of the nanorods we treated them as ellipsoids and used Gans' formula \cite{Gans12}, which is a standard approach in the literature, to obtain Equation~\eqref{eq:Gans}. However, since the synthesis rarely provides monodisperse nanorods, we account for their polydispersity by integrating over an aspect ratio distribution $f(X)$ (first term) as well as for the presence of an additional particle population, modelled as spheres\footnote{The spheres contribute to the AS spectrum a peak at approximately the same position as the TPP of the nanorods.} (second term), finally yielding for the absorbance $\gamma$:
\begin{align}
\label{eq:Gans}
\gamma &= \frac{2 \pi \epsilon_m^{3/2}}{3 \ln 10} \, \frac{l}{\lambda} \left [ \phi_R \int f(X') \text{d} X' \sum_{i=1}^3 \frac{\epsilon_ 2P_i^{-2}}{ \left( \epsilon_1 + \frac{1-P_i}{P_i} \epsilon_m \right)^2 + \epsilon_2^2} \right.\\ \nonumber
 &+ \left.27 \phi_S \frac{\epsilon_2}{(\epsilon_1+2\epsilon_m)^2 + \epsilon_2^2} \right ] \, ,
\end{align}
where $\lambda$ is the wavelength in vacuum, $l$ is the optical path, $\epsilon_m(\lambda)$ is the relative dielectric constant of the surrounding medium (water)\cite{Daimon:2007}, $\epsilon_1$ and $\epsilon_2$ are the real and imaginary parts of the dielectric constant of the particles $\epsilon(\lambda,R)$, derived from the (wavelength-dependent) value $\epsilon_{\text{bulk}}(\lambda)$ for bulk gold\footnote{See the ESI for a detailed discussion.} and corrected for the additional dissipation at the interface, which introduces the size dependence via the damping parameter $A$\cite{Hoevel:1993}:

\begin{align}
\label{eq:Hoevel}
\epsilon(\lambda,R) &= \epsilon_{\text{bulk}}(\lambda) + \frac{\omega_p^2}{\omega^2 + i \omega \gamma_0} \\ \nonumber
&- \frac{\omega_p^2}{\omega^2 + i \omega (\gamma_0 + A v_F/R)} \, ,
\end{align}
where $\omega = 2 \pi c/\lambda$ (with $c$ the speed of light in vacuum), $\omega_p = 2.183 \times 10^{15} \un{s^{-1}}$ the plasmon frequency, $\gamma_0 = 6.46 \times 10^{12} \un{s^{-1}}$ the bulk damping rate, $v_F = 1.4 \times 10^{6} \un{m/s}$ the Fermi velocity and $R$ stands for $R_S$, $R_R$ or $L/2$, depending on the contribution to be described.

The geometrical factors $P_i$ for a prolate ellipsoid along the major axis and along the transverse axes are respectively given by:

\begin{align}
\label{eq:Pi}
& P _1 =  \frac{1-e^2}{e^2} \left[ \frac{1}{2e} \ln \left( \frac{1+e}{1-e}\right) - 1\right],\\ \nonumber
& P_2 = P_3=\frac{1-P_1}{2} \, ,
\end{align}
and depend on the aspect ratio via the eccentricity $e = \sqrt{1-X'^{-2}}$. We assumed that the aspect ratio distribution $f(X')$ of the nanorods is given by a Schulz function \cite{Kotlarchyk:1983}, with average value $X$ and standard deviation $\sigma_{X}$. Each particle participates to the absorption \eqref{eq:Gans} in proportion to its volume, but $f(X')$ is not volume-weighted since the volume of the nanorods is not correlated with their aspect ratio. This feature has already been described in the literature \cite{Eustis:2006} and is fairly well verified by the analysis of our TEM data (see the ESI).

\subsection{SAXS experiments}
\label{subsec:SAXSexp}

The SAXS data was collected at the SWING beamline of the SOLEIL synchrotron (Saint-Aubin, France), at a beam energy of 12~keV and two sample-to-detector distances (1 and 6.5~m). The samples were contained in cylindrical glass capillaries (Mark-R\"{o}rchen) 1.5~mm in diameter. Preliminary data treatment (angular averaging and normalization) was done using the software \textsc{Foxtrot} developed at the beamline and yielded the intensity as a function of the scattering vector $I(q)$ in absolute units. Subsequent data fitting was done in \textsc{Igor Pro} using the NCNR SANS package \cite{Kline06}.

\subsection{TEM experiments}
\label{subsec:TEMexp}

The TEM was done using a Topcon instrument, at a voltage of $100\,\ce{kV}$ and a magnification of $14000$. The objects were identified automatically using the image treatment routines available in \textsc{Igor Pro}, individually validated by visual inspection and sorted into "nanospheres/nanocubes" and "nanorods" according to their aspect ratio. For each synthesis batch we analyzed at least 100 objects of each type.

\section{Results and discussion}
\label{sec:results}

\subsection{Absorbance spectroscopy}
\label{subsec:UVVis}

We obtain very good fits to the AS data using Equation~\eqref{eq:Gans}, with free parameters $ \phi_S$, $\phi_R$, $X$, $\sigma_{X}$ and $A$ (the radii are fixed at $R_S = 10 \, \text{nm}$ and $R_R = 5 \, \text{nm}$). The experimental data and the models are shown in Figure~\ref{fig:UV_Vis}.

\subsection{SAXS}
\label{subsec:SAXS}
The scattered intensity $I(q)$ is described by the sum of two contributions: spheres with polydisperse radius (mean value $R_S$ and variance $\sigma_S^2$) and cylinders (or rods) with a fixed length $L$ (the SAXS is not very sensitive to this parameter over the available $q$-range) and polydisperse radius (with mean value $R_R$ and variance $\sigma_R^2$). We assume that both $R_S$ and $R_R$ are Schulz-distributed \cite{Kotlarchyk:1983}. The scattering length densities for gold and water are: $\rho_{\ce{Au}} = 1.082 \times 10^{-4} \text{\AA}^{-2}$ and $\rho_{\ce{H_2O}} = 9.444 \times 10^{-6} \text{\AA}^{-2}$. \footnote{From http://www.ncnr.nist.gov/resources/activation/.} We performed the SAXS on concentrated solutions, with a total Au volume fraction $\phi \sim 10^{-4}$.

\begin{figure}[!h]
\includegraphics[width=0.45\textwidth,angle=0]{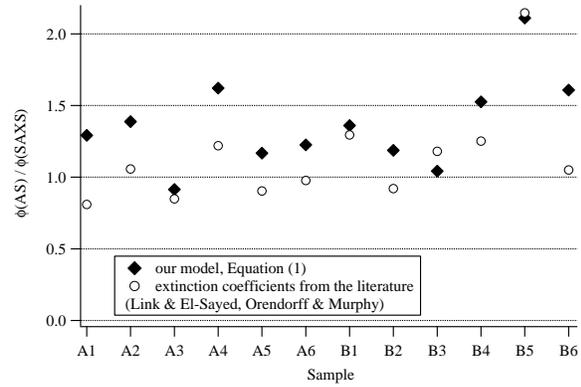}
\caption{Ratio of the two experimental estimates for the gold volume fraction $\phi (\text{AS})/\phi (\text{SAXS})$.  $\phi (\text{AS}) = \phi_R + \phi_S$ is obtained by fitting the absorption spectrum with Equation~\eqref{eq:Gans} (diamonds) and by using the atomic extinction coefficient from the literature (open dots); see the ESI for more details. $\phi (\text{SAXS})$ is obtained from the integrated SAXS intensity as described in the text.}
\label{fig:ratio}
\end{figure}

We started by estimating the volume fraction of gold particles in solution from the intensity integral over the entire reciprocal space. This SAXS ``invariant'' is given by: $Q = \int\limits_{0}^{\infty} \text{d}q \, q^2 I(q)$ and, for a two-phase system (in our case, gold and solvent) is related to their volume fractions by $Q = 2 \pi^2 (\rho_{\ce{Au}} - \rho_{\ce{H_2O}})^2 \phi (1 - \phi)$ \cite{Glatter:1982}. This treatment involves no assumption as to the particle morphology.


\begin{figure*}[t]
\centerline{\includegraphics[width=0.75\textwidth,angle=0]{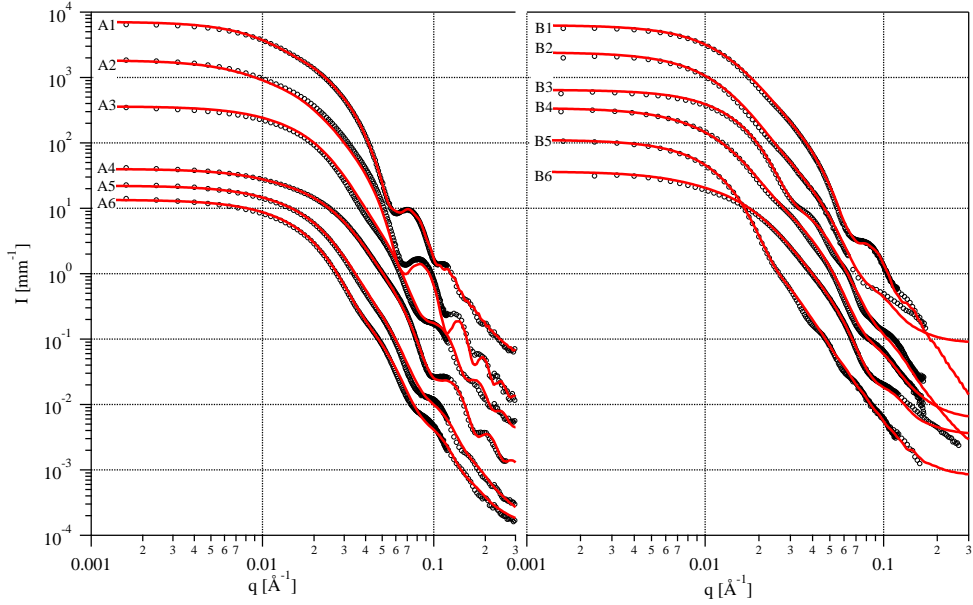}}
\caption{SAXS data (dots) and fits (solid lines). For clarity, the curves are shifted vertically by arbitrary amounts and only every fifth point is shown.}
\label{fig:SAXS}
\end{figure*}

The resulting gold volume fraction should be equal to that determined from fitting the absorption spectrum (accounting for the dilution ratio between the two samples).

In Figure~\ref{fig:ratio} we show the ratio $\phi (\text{AS})/\phi (\text{SAXS})$ for all studied samples (solid diamonds). Its average is about 1.25, an overestimation that should be kept in mind when using Equation~\eqref{eq:Gans} for determining the total particle concentration. We also use a different estimate for $\phi (\text{AS})$, based on literature results for the spheres\cite{Link:1999c,Mulvaney:1992} and the rods\cite{Orendorff:2006} obtained by inductively coupled plasma (ICP). In this case, the ratio $\phi (\text{AS})/\phi (\text{SAXS})$ (open dots) is remarkably close to 1, showing excellent agreement between ICP and SAXS. The detailed calculation is given in the ESI.

We then perform a full fit to the SAXS data (Figure~\ref{fig:SAXS}) with $\phi = \phi_R + \phi_S$ as obtained from the invariant estimation, while the ratio $\phi_S / \phi_R$ and the nanorod aspect ratio $X$ are fixed at the values extracted from the AS measurements. The fits are generally excellent, with only $(R_S, \sigma_S)$ and $(R_R, \sigma_R)$ as free parameters; see Table~\ref{table:all} for the values.

\begin{figure}
\includegraphics[width=0.5\textwidth,angle=0]{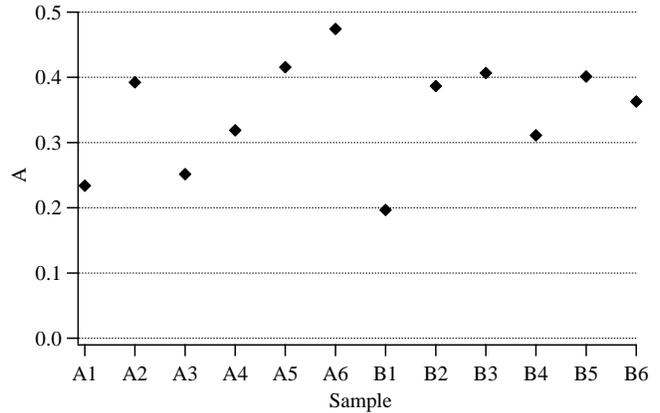}
\caption{Interface damping parameter $A$ defined in  Equation~\eqref{eq:Hoevel}. $A/R_S$ is determined from the fits to the AS data in Figure~\ref{fig:UV_Vis}, while $R_S$ is extracted from the SAXS data (Figure~\ref{fig:SAXS}).}
\label{fig:Damping}
\end{figure}

The AS model is only sensitive to the damping parameter $A$ in the combination $A/R_S$, via the corrected dielectric constant \eqref{eq:Hoevel} (The contribution of the transverse mode of the rods, which involves $A/R_R$, is much weaker than that of the spheres). Having determined $R_S$ independently from the SAXS data, we are now in a position to evaluate $A$ (see Figure~\ref{fig:Damping}). Its average value is about $A=0.35$, to be compared with the literature values of $0.6-2$ for gold spheres \cite{Quinten:1996,Alvarez:1997,Scaffardi:2006}.

\subsection{TEM}
\label{subsec:TEM}
The TEM images show the presence of two populations, separated by aspect ratio: elongated ($X \geq 2$) "nanorods" and isometric particles ($1 \leq X \leq 1.5$), which can be either rounded "nanospheres" or faceted "nanocubes". 

\begin{figure}
\includegraphics[width=0.45\textwidth,angle=0]{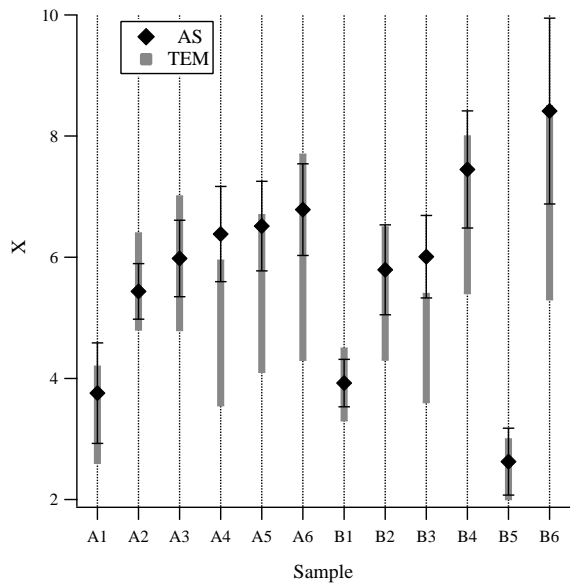}
\caption{Aspect ratio for the nanorods, as determined from the fits to the AS data in Figure~\ref{fig:UV_Vis} (diamonds) and from the TEM images (bars). The error bars (AS) and bar height (TEM) indicate the average value $\pm$ standard deviation: $X \pm \sigma _X$.}
\label{fig:AR}
\end{figure}

In Figure~\ref{fig:AR} we compare the aspect ratio $X$ and its relative dispersion $\sigma_X /X$ for the nanorods, determined from the AS data (diamonds) and from the TEM images (bars).

\begin{figure}
\includegraphics[width=0.45\textwidth,angle=0]{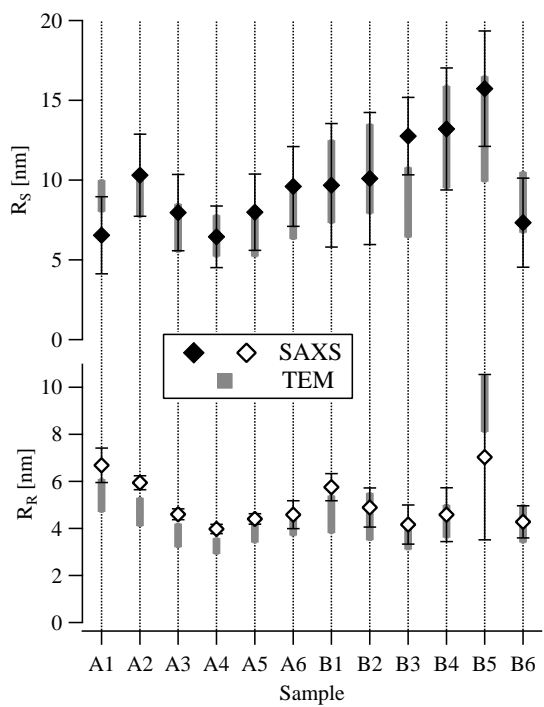}
\caption{Radii of the rods $R_R$ and spheres $R_S$ as determined from the fits to the SAXS data in Figure~\ref{fig:SAXS} (diamonds) and from the TEM images (shaded squares). The error bars (AS) and bar height (TEM) indicate the average value $\pm$ standard deviation: $R_S \pm \sigma _S$ and $R_R \pm \sigma _R$.}
\label{fig:Radii}
\end{figure}

Finally, in Figure~\ref{fig:Radii} we compare the values for the radii $R_R$ and $R_S$ of both particle populations, as obtained from SAXS and TEM. These graphs show that modelling the data obtained by the three different techniques provides values of the particle dimensions that are in good agreement. 

The results of the three techniques (AS, SAXS and TEM) are summarized in Table~\ref{table:all} for all samples.

\section{Simplified relations}
\label{sec:Calib}

Full-curve fitting of the AS data is done using a very fast and stable routine, implemented either in \textsc{MATLAB} (available on request from the first author, KS) or in \textsc{Igor Pro} (available on request from the last author, DC). It is nevertheless useful to extract simplified relations among the relevant parameters.

\begin{figure*} 
\centering
\includegraphics[width=0.4\textwidth,angle=0]{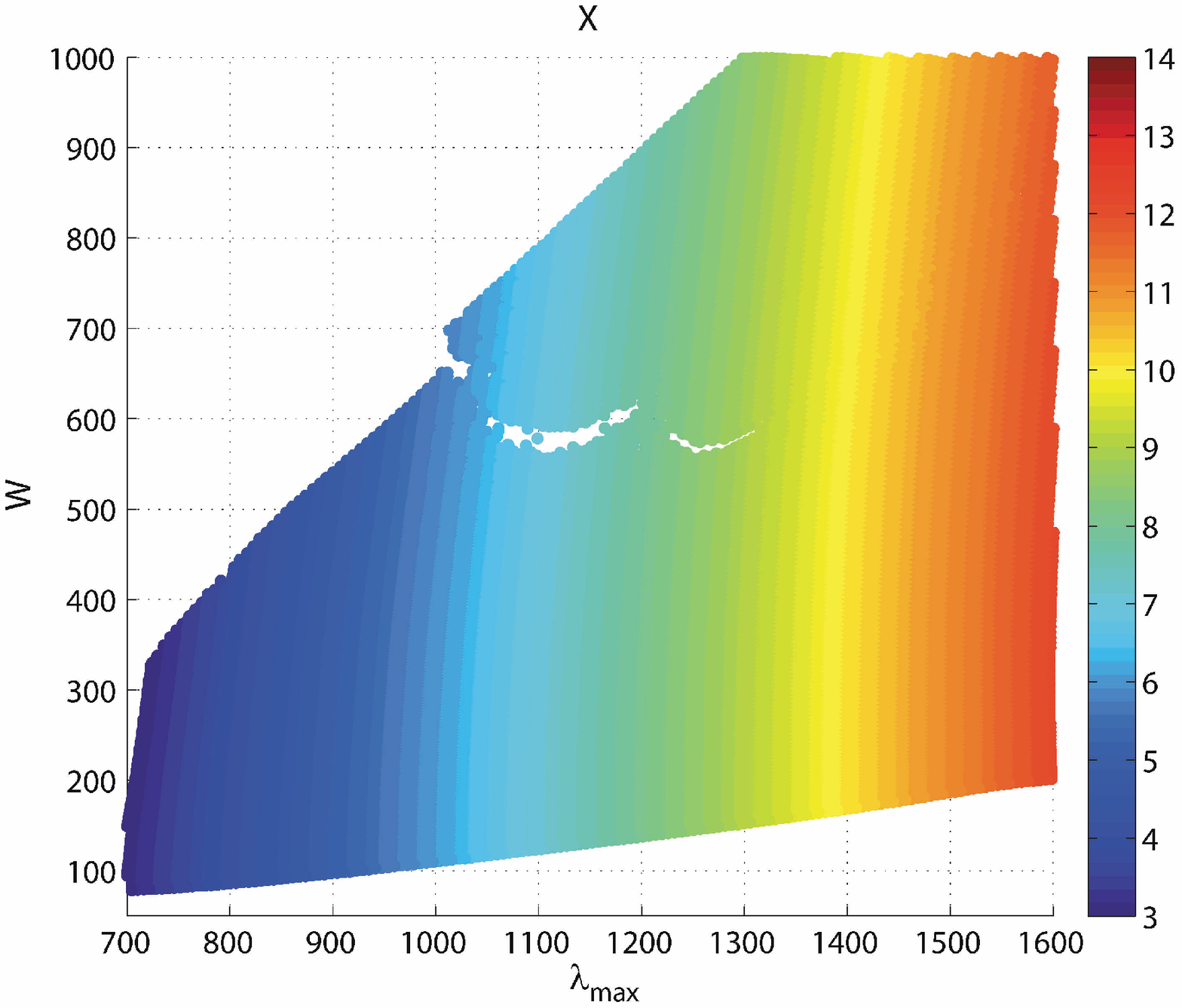}
\includegraphics[width=0.4\textwidth,angle=0]{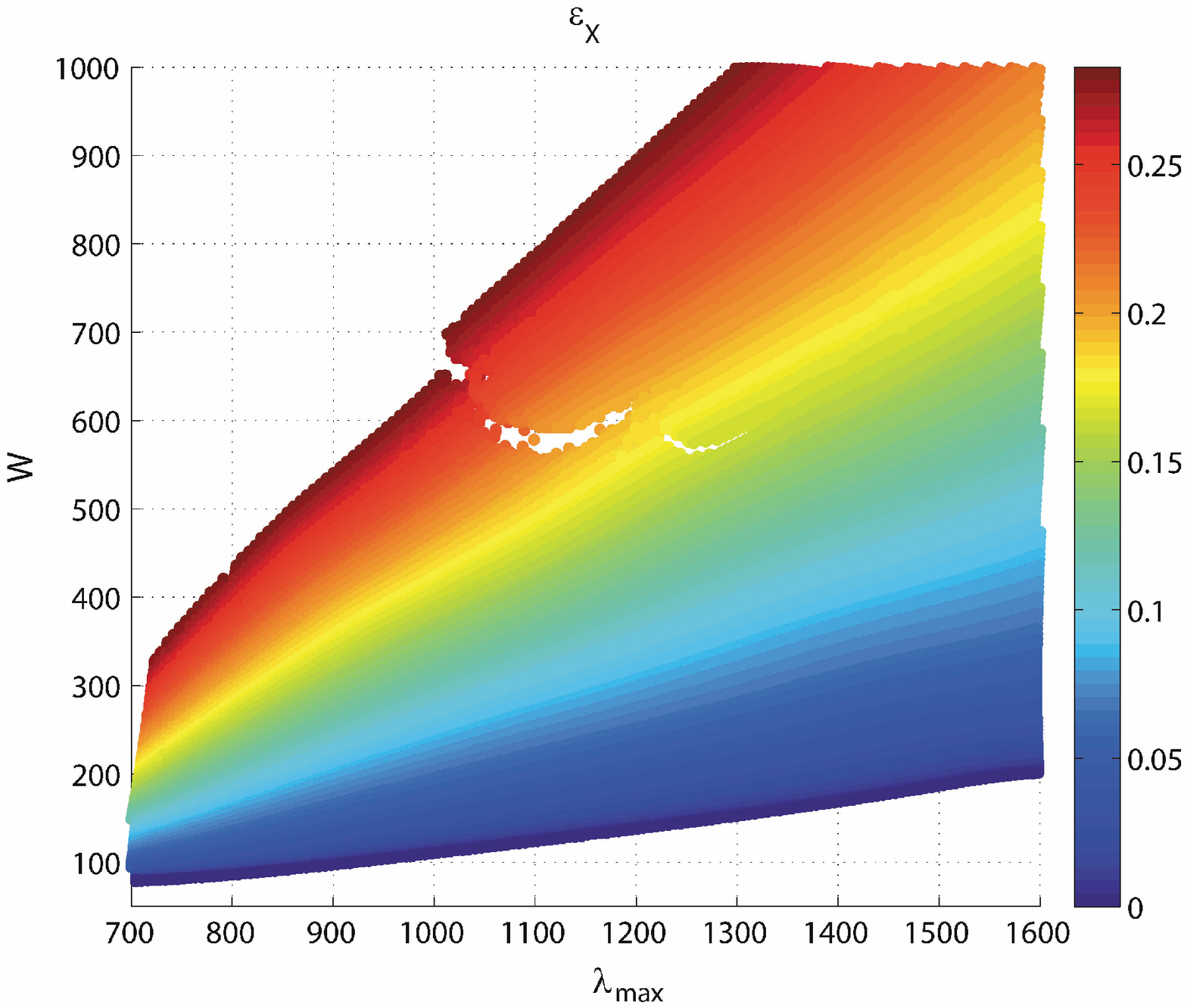}
\caption{Aspect ratio $X$ (left) and relative variance $\varepsilon_X$ (right) of the nanorod population simulated by Equation~\eqref{eq:Gans} in the main text as a function of $\lambda _{\text{max}}$ and $W$ (for $R_R/A = 14.3\,\text{nm}$) represented as color code.}
\label{fig:XS_LW}
\end{figure*}

From extensive simulations we thus obtained polynomial interpolations for the position and width of the LPP, $\lambda _{\text{max}}(X,\varepsilon_X)$ and $W(X,\varepsilon_X)$, as well as for the converse relations $X(\lambda _{\text{max}},W)$ and  $\varepsilon_X(\lambda _{\text{max}},W)$, where we introduced the relative polydispersity $\varepsilon_X = \sigma _X / X$. The latter relations, which are the most useful for evaluating the experimental results, are shown in Figure~\ref{fig:XS_LW}. If the parameters $\lambda _{\text{max}}$ and $W$ are known, the aspect ratio and its polydispersity can be read directly from the two graphs in the Figure.

The complete relations, as well as formulas for the concentration can be found in the ESI.

\begin{figure}[!h]
\includegraphics[width=0.5\textwidth,angle=0]{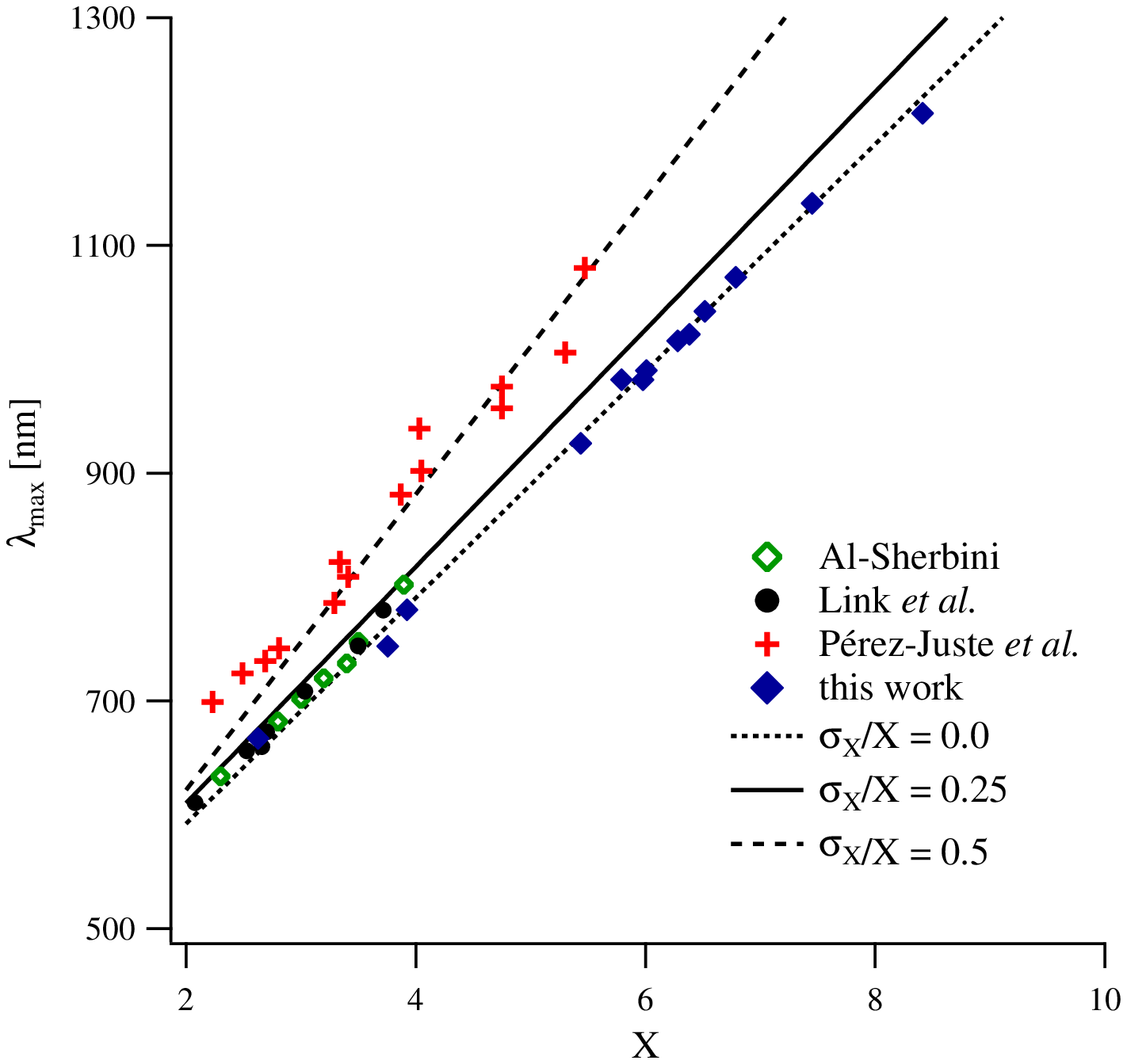}
\caption{Experimental data for the dependence of $\lambda _{\text{max}}$ on $X$, from several authors: Al-Sherbini \citep[(Figure~1b)]{AlSherbini:2004} ($\diamond$), Link \emph{et al.} \citep[(Figure~4, without reshaping)]{Link:1999} ($\bullet$), P\'{e}rez-Juste \emph{et al.} \citep[(Tables~S1 and S2)]{Perez:2005} ($+$) and the present work ($\filleddiamond$). Predictions of the simplified model (see the ESI), for various relative polydispersities $\varepsilon_X = \sigma _X/X=0$ (dotted line), 0.25 (solid line) and 0.5 (dashed line). For all curves, $R_R/A = 14.3 \un{nm}$.}
\label{fig:Final_comp}
\end{figure}

\section{The importance of polydispersity}
\label{sec:Poly}

In Figure~\ref{fig:Final_comp} we compare the simplified model discussed above to the data in this paper and to results published by other groups for the dependence of $\lambda _{\text{max}}$ on $X$.

For perfectly monodisperse systems, the dependence $\lambda _{\text{max}}(X,0)$ (dotted line in Figure~\ref{fig:Final_comp}) is very close to that given by Yan \emph{et al.}\cite{Yan:2003}. However, a quantitative description of the experimental data requires accounting for the polydispersity: $\varepsilon_X \leq 0.15$ for our experimental data (in good agreement with the values presented in Table~\ref{table:all}), $\varepsilon_X \simeq 0.15$ for those of \citet{AlSherbini:2004} and \citet{Link:1999} and $\varepsilon_X \simeq 0.5$ for \citet{Perez:2005}. It is obvious from Figure~\ref{fig:Final_comp} that a unique curve $\lambda _{\text{max}}(X)$ cannot describe all the experimental data. We emphasize that the polydispersity correction is very significant: for instance, $\lambda _{\text{max}}(7,0.5)- \lambda _{\text{max}}(7,0.0) \simeq 180 \, \, \text{nm}$!

\section{Conclusions}
\label{sec:Conc}

\begin{figure*}[htbp]
\centerline{\includegraphics[width=0.9\textwidth,angle=0]{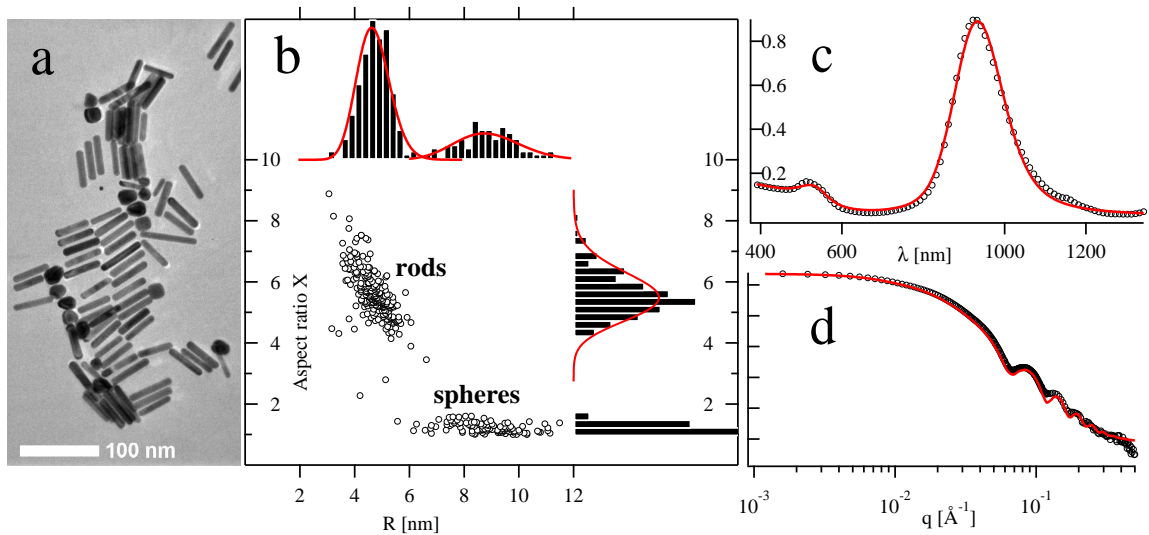}}
\caption{The three techniques at a glance for sample A2. (a) Selection from a TEM image. (b) TEM analysis: aspect ratio vs. radius for all analyzed particles, with histograms of both parameters. The solid lines are Schulz distributions with the average values and standard deviations given in Table~\ref{table:all}. (c) AS data from Figure~\ref{fig:UV_Vis}. (d) SAXS data from Figure~\ref{fig:SAXS}.}
\label{fig:Fig_A2}
\end{figure*}

We characterized solutions of gold nanoparticles using three different techniques, summarized in Figure~\ref{fig:Fig_A2}.

Our goal was to show that fitting the full AS spectrum allows retrieving detailed information about particle dimensions and their polydispersity in mixtures of gold nanospheres and nanorods. After comparison with the SAXS and TEM measurements and with other results in the literature we conclude that:

\begin{itemize}
\item The total volume fraction of particles $\phi = \phi_R + \phi_S$ inferred from the AS data overestimates by about 25\% the SAXS results (which we consider as reliable).
\item The aspect ratio of the nanorods and its dispersion $(X, \sigma_X)$ are determined accurately by AS and validated by a comparison with the TEM data.
\item The radii for both the rods and spheres $(R_R, R_S)$ are measured by SAXS and confirmed by the TEM data. Combining $X$ and $R_R$ then yields a reliable estimate for the length $L$, which is not directly accessible via either AS or SAXS alone.
\item The model for $\lambda _{\text{max}}(X,\sigma _X)$ successfully describes the experimental results from various research groups; in particular, the polydispersity $\sigma _X$ plays an important role.
\end{itemize}

From the AS model, we extracted polynomial approximations for $X$, $\sigma _X$, $\phi_R$ and $\phi_S$ as a function of the features of the AS spectrum (height, position and width of the longitudinal peak and height of the transverse peak). These formulas can be easily implemented in any data treatment software and should therefore provide a fast and simple way of characterizing mixtures of gold nanoparticles, both \textit{ex situ}, but also during synthesis, to characterize the kinetics of particle formation and to optimize the preparation method.

Some points we have not considered in the present paper, but that would deserve development in future work are:
\begin{itemize}
\item The role of the finite size of the particles with respect to the wavelength. With increasing size, the electrostatic approximation becomes less and less accurate. Corrections are then needed to describe the position and amplitude of the plasmon peaks.
\item The influence of the organic layer at the interface between the particle and the medium. The nature of the ligands adsorbed onto the surface of the nanoparticles modifies their optical response: for homogeneous molecular layers, this effect might be described by a coated ellipsoid model.
\item Extending the approach to other particle shapes, either rounded (dumbbells, dogbones) or faceted (cubes, prisms, plates). Changing the synthesis parameters leads to a wide variety of shapes, each of them exhibiting interesting features and potential applications. It would thus be extremely useful to extend the model in this direction.
\end{itemize}

\section*{Acknowledgements}
We acknowledge the Triangle de la Physique for financial support (project 2011-083T) and the SOLEIL synchrotron for the provision of synchrotron radiation facilities (Proposal ID 20120731) and we would like to thank Javier Perez for assistance in using beamline SWING. We thank A. Lecchi for help with setting up the nanoparticle synthesis.

\bibliographystyle{rsc}
\bibliography{metamat}

\end{document}